\documentclass[a4paper,12pt]{article}
\pdfoutput=1
\usepackage{graphicx}
\usepackage{mathtools}
\usepackage{amssymb}
\usepackage{amsfonts}
\usepackage[footnotesize]{caption}
\usepackage[font=scriptsize]{subcaption}
\usepackage{color}
\usepackage{braket}
\usepackage{cite}
\usepackage{hyperref}
\usepackage{url}
\usepackage{multirow}
\usepackage{relsize}
\usepackage{fullpage}
\usepackage{makecell}
\usepackage{blkarray}
\usepackage{fullpage}
\usepackage{bbm,bbold}

\usepackage{url}

\usepackage[utf8]{inputenc}

\newcommand{\newc}{\newcommand}
\newc{\be}{\begin{equation}}
\newc{\ee}{\end{equation}}
\newc{\bea}{\begin{eqnarray}}
\newc{\eea}{\end{eqnarray}}
\newc{\simlt}{~\mbox{\smaller\(\lesssim\)}~}
\newc{\simgt}{~\mbox{\smaller\(\gtrsim\)}~}

\allowdisplaybreaks

\begin{document}

\begin{titlepage}
\begin{center}
{\bf\Large
\boldmath{
Trimaximal TM$_1$ mixing with two modular $S_4$ groups
}
} 
\\[12mm]
Stephen~F.~King$^{\star}$%
\footnote{E-mail: \texttt{king@soton.ac.uk}},
Ye-Ling~Zhou$^{\star}$%
\footnote{E-mail: \texttt{ye-ling.zhou@soton.ac.uk}},
\\[-2mm]
\end{center}
\vspace*{0.50cm}
\centerline{$^{\star}$ \it
School of Physics and Astronomy, University of Southampton,}
\centerline{\it
Southampton SO17 1BJ, United Kingdom }

\vspace*{1.20cm}

\begin{abstract}
{\noindent
We discuss a minimal flavour model with 
twin modular symmetries, leading to trimaximal TM$_1$ lepton mixing in which the 
first column of the tri-bimaximal lepton mixing matrix is preserved.
The model involves two modular $S_4$ groups, one acting in the neutrino sector, associated with a modulus field value $\tau_{SU}$ with residual $Z^{SU}_2$ symmetry,
and one acting in the charged lepton sector, associated with a modulus field value $\tau_{T}$ with residual $Z^{T}_3$ symmetry. 
Apart from the predictions of TM$_1$ mixing, the model leads to a new neutrino mass sum rule which implies lower bounds on neutrino masses close to current limits from 
neutrinoless double beta decay experiments and cosmology.
}
\end{abstract}
\end{titlepage}

\section{Introduction}
The discovery of neutrino masses and lepton mixing opened up a new direction in physics beyond Standard Model (SM) focussed on understanding their theoretical origin.
An elegant possibility remains 
the classical type-Ia seesaw mechanism \cite{Minkowski:1977sc, Yanagida:1979ss, Gell-Mann:1979ss, Glashow:1979ss, Mohapatra:1979ia, Schechter:1980gr, Schechter:1981cv} involving right-handed neutrinos, which, after being integrated out,
yield the Weinberg operators $H_uH_uL_iL_j$ with $H_u=H$ being the SM Higgs doublet and $L_i$
 a lepton doublet of the $i$th flavour~\footnote{
An alternative type-Ib seesaw mechanism,
yielding the new Weinberg operators $H_u\tilde{H}_dL_iL_j$ with 
$\tilde{H}_d$ being a charge conjugated second Higgs doublet with opposite hypercharge, was proposed in~\cite{Hernandez-Garcia:2019uof} recently.}.
To explain the observed approximate tri-bimaximal (TBM)
lepton mixing, one has to go beyond the seesaw mechanism and consider to
impose a non-Abelian discrete flavour symmetry~\cite{King:2013eh,King:2017guk}.
For example, $S_4$ can be used to account for trimaximal TM$_1$ mixing
\cite{Varzielas:2012pa, Luhn:2013vna}, which is imposed by a residual $Z^{SU}_2$ symmetry in the neutrino sector and a residual
$Z_3^T$ symmetry in the charged lepton sector~\footnote{We apply the standard convention of the 
$S_4$ generators $S$, $T$, and $U$
where $S^2=T^3=U^2=(ST)^3=(SU)^2=(TU)^2=I$~\cite{King:2013eh} hold.}.
However all existing realistic models typically involve several flavon fields with non-trivial vacuum alignments.

Non-Abelian discrete flavour symmetries have been widely used in models of lepton flavour mixing for decades, but the nature of non-Abelian discrete flavour symmetry is still unclear. It might be an effective remnant symmetry 
after a continuous non-Abelian symmetry breaking
\cite{deMedeirosVarzielas:2005qg, Koide:2007sr, Banks:2010zn, Luhn:2011ip, Merle:2011vy, Wu:2012ria, Rachlin:2017rvm, King:2018fke}, 
or a fundamental symmetry of spacetime in extra dimensions \cite{Asaka:2001eh, Altarelli:2006kg, Kobayashi:2006wq, Altarelli:2008bg, Adulpravitchai:2009id, Burrows:2009pi, Adulpravitchai:2010na, Burrows:2010wz, deAnda:2018oik, Kobayashi:2018rad, deAnda:2018yfp, Baur:2019kwi}. 
In the latter case, a non-Abelian discrete symmetry  could either arise as an accidental symmetry
of orbifolding
(see  \cite{Kobayashi:2008ih,deAnda:2018oik,Olguin-Trejo:2018wpw, Mutter:2018sra} for recent discussion with two extra dimensions) or as a subgroup of the so-called modular symmetry. The modular symmetry \cite{Giveon:1988tt} is an infinite symmetry of the extra dimensional lattice arising from superstring theory \cite{Ferrara:1989bc,Ferrara:1989qb}
\footnote{Recently, the geometric connection between the origin of the flavour symmetry due to 
modular symmetry and that due to orbifolding with two extra dimensions
has been discussed, e.g., in \cite{deAnda:2018ecu, Kobayashi:2018bff}.}.
Indeed, it has been suggested that a finite subgroup of the modular group, when interpreted as a flavour symmetry, might be helpful for an explanation for lepton mixing
\cite{Altarelli:2005yx, deAdelhartToorop:2011re, Kobayashi:2019mna}.

Recently such a finite modular symmetry has been proposed as the direct origin of flavour mixing. In this approach, Yukawa and mass textures arise not from flavon fields, but modular forms with even modular weights which are holomorphic functions of a modulus field 
\cite{Feruglio:2017spp} \footnote{Very recently, this approach has been extended to include odd weight modular forms \cite{Liu:2019khw}. }. The complex modulus field 
$\tau$ acquires a vacuum expectation value (VEV) and eventually determines the flavour structure. 
The finite modular groups $\Gamma_2\simeq S_3$~\cite{Kobayashi:2018vbk, Kobayashi:2018wkl}, 
$\Gamma_3\simeq A_4$~\cite{Feruglio:2017spp, Criado:2018thu, Kobayashi:2018scp, Okada:2018yrn, Kobayashi:2018wkl, Novichkov:2018yse, Ding:2019zxk}, 
$\Gamma_4\simeq S_4$~\cite{Penedo:2018nmg, Novichkov:2018ovf} 
and 
$\Gamma_5\simeq A_5$~\cite{Novichkov:2018nkm, Ding:2019xna}
have been considered, in which special Yukawa textures are consequences of the modular forms. Compared with the framework of traditional flavour model constructions, only a minimal set of flavons (or no flavons at all) 
need to be introduced in this framework~\footnote{Extensions to flavour mixing in the quark sector are given in \cite{Kobayashi:2018wkl,Okada:2018yrn, Okada:2019uoy, Kobayashi:2019rzp}.}, making such an approach very attractive.

For flavour models with finite modular symmetry outlined above, only one single modulus field $\tau$ is usually included,
corresponding to a single finite modular group symmetry $\Gamma_N$.
It has been pointed out that particular modular forms at some special values of 
the modulus VEV
preserve a residual subgroup of the finite modular symmetry.
Such an idea was discussed in \cite{Novichkov:2018yse} where residual symmetries are considered as subgroups of the modular $A_4$ symmetry. Making use of two moduli fields with VEVs preserving different residual symmetries, i.e., $Z_3$ in the charged lepton sector and $Z_2$ in the neutrino sector, it was shown how 
trimaximal TM$_2$ mixing might be realised~\cite{Novichkov:2018yse}. A brief discussion of residual symmetry after the breaking of modular $S_4$ symmetry has  also been given in \cite{Novichkov:2018ovf}.

In a recent paper \cite{deMedeirosVarzielas:2019cyj}, two of us extended the formalism of finite modular symmetry to the case of multiple
moduli fields $\tau_J$ ($J=1, \ldots M$) associated with the finite modular symmetry
$\Gamma_{N_1}^1\times \Gamma_{N_2}^2 \times \cdots \times \Gamma_{N_M}^M$.
This is motivated by superstring theory 
which involves six compact extra dimensions, suggesting the introduction of three modular symmetries associated with three
different factorised tori in the simplest compactifications.
As an example, we presented the first consistent 
example of a flavour model of leptons with multiple modular $S_4$ symmetries
interpreted as a flavour symmetry.
The considered model involved three finite modular symmetries $S_4^A$, $S_4^B$
and $S_4^C$, associated with two right-handed neutrinos and the charged lepton sector, respectively,
broken by two bi-triplet scalars to their diagonal subgroup. The low energy effective theory consisted
of a single $S_4$ modular symmetry with three independent modular fields $\tau_A$, $\tau_B$ and $\tau_C$,
which preserve the residual modular subgroups $Z_3^A$, $Z_2^B$ and $Z_3^C$, 
in their respective sectors
leading to trimaximal TM$_1$ lepton mixing, in which the first column of the tri-bimaximal mixing matrix is achieved,
in excellent agreement with current data, without requiring any flavons.

In the present paper we discuss a simpler
model of TM$_1$ lepton mixing via two modular $S_4$ groups, one $S_4^{\nu}$ acting in the neutrino sector, associated with a modulus field value $\tau_{SU}$ with residual $Z^{SU}_2$ symmetry,
and one $S_4^{l}$ acting in the charged lepton sector, associated with a modulus field value $\tau_{T}$ with residual $Z^{T}_3$ symmetry. 
The two moduli fields are assumed to be ``stabilised" at these symmetric points, and there are no other flavons, making the model very economical and predictive. In particular it leads to a new neutrino mass sum rule which implies sizeable neutrino masses sensitive to neutrinoless double beta decay and cosmological probes. The main difference between the present model and the one in \cite{deMedeirosVarzielas:2019cyj}, is that here we assume that there are three right-handed neutrinos in a triplet of an $S_4$, whereas the previous model assumed two right-handed neutrinos which were $S_4$ singlets. The resulting model here is very similar to the ``semi-direct'' models of traditional flavour symmetry. However the predictions are different due to the smaller number of parameters, leading to a new and testable neutrino mass sum rule.

The rest of the paper is organised in the following.
In section~\ref{S4} we first focus on the case of the single finite modular $S_4$ symmetry, with residual symmetry
arising from the moduli stabilisers. We then generalise the results to the case of two modular $S_4$ groups.
In section~\ref{model} we propose a model based on $S_4^{\nu}\times S_4^{l}$ with 
two moduli fields, which is broken to a single diagonal $S_4$ with two independent moduli fields at low energies, whose stabilisers lead to different remnant symmetry in the different sectors, which may be used to enforce 
trimaximal TM$_1$ mixing with the new neutrino mass sum rule.
Section~\ref{conclusion} concludes the paper.

\section{$S_4$ modular symmetries}
\label{S4}
\subsection{A single $S_4$ modular group}

The modular group $\overline{\Gamma}$ acting on the modulus field $\tau$ as linear fractional transformations
\begin{eqnarray} \label{eq:modular_transformation}
\gamma:~ 
\tau \to \gamma \tau = \frac{a \tau + b}{c \tau + d}\,,
\end{eqnarray}
where the modulus field $\tau$ is defined on the upper complex plane ${\rm Im}(\tau)>0$,  $a$, $b$, $c$, and $d$ are integers and satisfy $ad-bc=1$.
It is convenient to represent each element of $\overline{\Gamma}$ by a two by two matrix \footnote{Note that it may not be a unitary matrix.}. In this way, $\overline{\Gamma}$ is expressed as 
\begin{eqnarray}
\overline{\Gamma} = \left\{ \begin{pmatrix} a & b \\ c & d \end{pmatrix} / (\pm \mathbf{1})\,,~ a, b, c, d \in \mathbb{Z}, ~~  ad-bc=1  \right\} \,.
\end{eqnarray}
The modular group is isomorphic to the projective spatial linear group $PSL(2,\mathbb{Z}) = SL(2,\mathbb{Z})/\mathbb{Z}_2$. 
It has two generators, $S_\tau$ and $T_\tau$, satisfying $S_\tau^2 = (S_\tau T_\tau)^3 = \mathbf{1}$. These generators act on the modulus $\tau$ in the following way, 
\begin{eqnarray}
S_\tau: \tau \to -\frac{1}{\tau} \,, \quad
T_\tau: \tau \to \tau + 1\,,
\end{eqnarray}
respectively. Representing them by two by two matrices, we obtain 
\begin{eqnarray}
S_\tau=\begin{pmatrix} 0 & 1 \\ -1 & 0 \end{pmatrix}\,, \hspace{1cm}
T_\tau=\begin{pmatrix} 1 & 1 \\ 0 & 1 \end{pmatrix} \,.
\end{eqnarray}
$\overline{\Gamma}$ is a discrete but infinite group. By requiring $a, d = 1~({\rm mod}~4)$ and $b, c =  0~({\rm mod}~4)$, i.e., 
\begin{eqnarray} \label{eq:mode_N}
a = 4 k_a+1\,,~ 
d = 4 k_d +1\,,~
b = 4 k_b\,, ~
c = 4 k_c\,,
\end{eqnarray}
where $k_a$, $k_b$, $k_c$ and $k_d$ are all integers, we obtain a subset of $\overline{\Gamma}$ labelled as 
\begin{eqnarray}
\overline{\Gamma}(4) = \left\{ \begin{pmatrix} a & b \\ c & d \end{pmatrix} \in PSL(2,\mathbb{Z}), ~~  \begin{pmatrix} a & b \\ c & d \end{pmatrix} = \begin{pmatrix} 1 & 0 \\ 0 & 1 \end{pmatrix} ~~ ({\rm mod}~ 4) \right\} \,.
\end{eqnarray}
It is also an infinite group. 
The quotient group $\Gamma_4 = \overline{\Gamma}/\overline{\Gamma}(4)$ is a finite modular group. 
It is equivalently obtained by imposing $T_\tau^4 = \mathbf{1}$. As $\Gamma_4$ is a subgroup of $\overline{\Gamma}$, its elements can also be represented as two by two matrices, but the representation matrices are not unique. Since $S_4$ is the quotient group $\overline{\Gamma}/\overline{\Gamma}(4)$, with the help of Eq.~\eqref{eq:mode_N}, we know that any element $\gamma$ of $S_4$ which can be written as
\begin{eqnarray} \label{eq:rep_1}
\begin{pmatrix} a & b \\ c & d \end{pmatrix}
\end{eqnarray} 
is identical to be represented in the form
\begin{eqnarray} \label{eq:rep_2}
\eta \begin{pmatrix} 4 k_a +a & 4 k_b +b \\ 4 k_c +c & 4 k_d +d \end{pmatrix} \,,
\end{eqnarray} 
where the integers $k_a$, $k_b$, $k_c$ and $k_d$ satisfy $4 k_a k_d + a k_d + d k_a = 4 k_b k_c + b k_c + c k_b$ and $\eta= \pm 1$. This is just a mathematical redundancy. Selecting a different two by two representation matrix gives no physical difference. 

The finite modular group $\Gamma_4$ is isomorphic to $S_4$,  the permutation group of four objects. 
In other word, $S_\tau$ and $T_\tau$ which satisfy $S_\tau^2  = (S_\tau T_\tau)^3 = T_\tau^4 =1$, can be used as generators of $S_4$. 
In the literature of flavour symmetry studies, it is more popular to use a different set of generators, $S$, $T$ and $U$, which satisfy $S^2 = T^3 = U^2 = (ST)^3 = (SU)^2 = (TU)^2 = 1$, to generate $S_4$. These generators can be represented by $S_\tau$ and $T_\tau$ as
\begin{eqnarray}
T = S_\tau T_\tau \,,\qquad
S = T_\tau^2 \,,\qquad
U = T_\tau S_\tau T_\tau^2 S_\tau \,.
\end{eqnarray}
With the requirement $\tau = \tau +4$, $S$, $T$ and $U$ can be represented by two by two matrices such as
\begin{eqnarray} \label{eq:STU}
T=\begin{pmatrix} 0 & 1 \\ -1 & -1 \end{pmatrix} \,, \quad
S=\begin{pmatrix} 1 & 2 \\ 0 & 1 \end{pmatrix}\,, \quad
U=\begin{pmatrix} 1 & -1 \\ 2 & -1 \end{pmatrix} \,. 
\end{eqnarray}
Again, we mention that representation matrices of these elements are not unique. Different representation matrices are obtained by considering the correlation between Eqs.~\eqref{eq:rep_1} and \eqref{eq:rep_2}. 
We also list a two by two matrix for $SU = S_\tau T_\tau S_\tau T_\tau^{-1} S_\tau$ \footnote{The product $SU$ gives 
\begin{eqnarray*} 
\begin{pmatrix}5 & -3 \\ 2 & -1 \end{pmatrix} = (-1) \begin{pmatrix}4 \times (-1)-1 & 4 \times 1-1 \\ 4 \times (-1)+2 & 4 \times 0+1 \end{pmatrix} \,.
\end{eqnarray*} 
Applying Eq.~\eqref{eq:rep_2}, we arrive at Eq.~\eqref{eq:SU}.}
\begin{eqnarray} \label{eq:SU}
SU=\begin{pmatrix} -1 & -1 \\ 2 & 1 \end{pmatrix}\,.
\end{eqnarray}
This generator is important for the trimaximal TM$_1$ mixing in the classical flavour model building (see, e.g., \cite{Luhn:2013vna}) and will also be used for our model construction next section.

In the framework of $\mathcal{N}=1$ supersymmetry with the $S_4$ modular symmetry, 
the superpotential $W(\phi_i;\tau)$ is in general a function of the modulus field $\tau$ and superfields $\phi_i$. Under the modular transformation, the superpotential should be invariant \cite{Ferrara:1989bc}. 
Expanding the superpotential $W(\phi_i;\tau)$ in powers of the superfields $\phi_i$, we obtain
\begin{eqnarray}
W(\phi_i;\tau) = \sum_n \sum_{\{i_1, \cdots, i_n\}} \sum_{I_Y} \left( Y_{I_Y} \phi_{i_1} \cdots \phi_{i_n} \right)_{\mathbf{1}} \,,
\end{eqnarray}
where $Y_{I_Y}$ represents a collection of coefficients of the couplings. The chiral superfield  $\phi_i$, as a function of $\tau$ (but does not need to be a modular form), transforms as \cite{Ferrara:1989bc},
\begin{eqnarray}
 \phi_i(\tau) \to \phi_i(\gamma\tau) = (c\tau + d)^{-2k_i} \rho_{I_i}(\gamma) \phi_i(\tau)\,,
 \label{eq:field_transformation}
\end{eqnarray}
where $-2k_i$ (with $k_i$ being an integer) is the modular weight of $\phi_i$, ${I_i}$ denotes the representation of $\phi_i$ and $\rho_{I_i}(\gamma)$ is a unitary representation matrix of $\gamma$ with $\gamma \in S_4$. 
The coefficients  $Y_{I_Y}$ transform as a multiplet modular form of weight $2k_Y$ and with the representation $I_Y$, 
\begin{eqnarray} \label{eq:form_transformation}
Y_{I_Y}(\tau) \to Y_{I_Y}(\gamma \tau) = (c\tau + d)^{2k_Y} \rho_{I_Y}(\gamma) Y_{I_Y}(\tau) \,,
\end{eqnarray}
where $k_Y = k_{i_1} + \cdots + k_{i_n}$ is required to be a non-negative integer. The representation and weight of $Y_{I_Y}$ are constrained due to the invariance of the operator under the $S_4$ modular transformation. For $k_Y=1$, there are 5 modular forms $Y_i(\tau)$ for $i=1,2,3,4,5$, which form a doublet $\mathbf{2}$ and a triplet $\mathbf{3}'$ of $S_4$, 
\begin{eqnarray} 
Y^{(2)}_{\mathbf{2}}(\tau) = \begin{pmatrix} Y_1(\tau) \\ Y_2(\tau) \end{pmatrix} \,,\quad
Y^{(2)}_{\mathbf{3}'}(\tau) = \begin{pmatrix} Y_3(\tau) \\ Y_4(\tau) \\ Y_5(\tau) \end{pmatrix} \,.
\end{eqnarray}
Specifically, an algebra between $Y_3$, $Y_4$ and $Y_5$
\begin{eqnarray}  \label{eq:algebra}
(Y_3^2 + 2 Y_4 Y_5)^2 = (Y_4^2 + 2 Y_3 Y_5) (Y_5^2 + 2 Y_3 Y_4)
\end{eqnarray}
is satisfied \cite{Penedo:2018nmg}. This constraint is independent of the value of $\tau$, and essential to cover the modular space of $\Gamma_4$. Contracting these modular forms gives rise to modular forms with weights $2k_Y=4$, 
\begin{eqnarray} 
Y^{(4)}_{\mathbf{1}}(\tau) = Y_1Y_2 \,,&&
Y^{(4)}_{\mathbf{2}}(\tau) = \begin{pmatrix} Y_2^2 \\ Y_1^2 \end{pmatrix}\,, \nonumber\\
Y^{(4)}_{\mathbf{3}}(\tau) = \begin{pmatrix} Y_1 Y_4 - Y_2 Y_5 \\ Y_1 Y_5 - Y_2 Y_4 \\ Y_1 Y_3 - Y_2 Y_4 \end{pmatrix}\,,&&
Y^{(4)}_{\mathbf{3}'}(\tau) = \begin{pmatrix} Y_1 Y_4 + Y_2 Y_5 \\ Y_1 Y_5 + Y_2 Y_4 \\ Y_1 Y_3 + Y_2 Y_4 \end{pmatrix} \,.
\end{eqnarray}
Modular forms with higher weights can all be constructed from $Y_i$. We refer to  \cite{Penedo:2018nmg} for detailed discussions. 


It is helpful to summarise the special properties of stabilisers and their relations with residual modular symmetries. We gave a thorough discussion on this issue in \cite{deMedeirosVarzielas:2019cyj}. Here we will mention four stabilisers which are relevant to the current work,  
\begin{eqnarray}
\tau_T= \omega = - \frac{1}{2} + i \frac{\sqrt{3}}{2} \,,~
\tau_S= i\infty \,,~
\tau_U=\frac{1}{2} + \frac{i}{2} \,,~
\tau_{SU}=-\frac{1}{2} + \frac{i}{2} \,.
\end{eqnarray}
Given any element $\gamma$ in a modular group, 
a stabiliser of $\gamma$ is a special value of the modulus field, denoted as $\tau_\gamma$, which satisfies $\gamma \tau_\gamma = \tau_\gamma$. 
If the modulus $\tau$ gains a VEV at the stabiliser, $\langle \tau \rangle = \tau_\gamma$, an Abelian residual modular symmetry generated by $\gamma$ is preserved. Specifically, for $\langle \tau \rangle = \tau_T, \tau_S, \tau_U, \tau_{SU}$, residual symmetries $Z_3^T$, $Z_2^S$, $Z_2^U$ and $Z_2^{SU}$ are preserved, respectively \footnote{The stabiliser of an element $\gamma$ may not be unique. We will not discuss other stabilisers that preserve $Z_3^T$, $Z_2^S$, $Z_2^U$ or $Z_2^{SU}$. }.

A modular form at a stabiliser takes an interesting weight-dependent direction. Starting from $Y_I(\gamma \tau_\gamma) = Y_I(\tau_\gamma)$ and following the standard transformation property in Eq. \eqref{eq:form_transformation}, one arrives at 
\begin{eqnarray} \label{eq:yukawa_eigenvector}
\rho_I(\gamma) Y_I(\tau_\gamma) = (c\tau_\gamma + d)^{-2k} Y_I(\tau_\gamma) \,. 
\end{eqnarray} 
Therefore, a modular form at a stabiliser $Y_I(\tau_\gamma)$ is an eigenvector of the representation matrix $\rho_I(\gamma)$ with respective eigenvalue $(c\tau_\gamma + d)^{-2k}$. 
If $(c\tau_\gamma + d)^{-2k}=1$ is satisfied, $\rho_I(\gamma) Y_I(\tau_\gamma) = Y_I(\tau_\gamma)$, the residual modular symmetry is reduced to the residual flavour symmetry. 
Otherwise, the residual modular symmetry is different from a residual flavour symmetry. 

We consider triplet modular forms $Y_{\mathbf{3}^{(\prime_\gamma)}}^{(2k)}(\tau)$ at $\tau_\gamma = \tau_S$, $\tau_T$, $\tau_U$ and $\tau_{SU}$. The eigenvalue $(c\tau_\gamma+d)^{-2k}$ at these stabilisers is respectively given by 
\begin{eqnarray}
(c\tau_\gamma+d)^{-2k} = \left\{ \begin{array}{ll} 
(0\tau_S+1)^{-2k} \equiv 1 & \text{for } \gamma=S \text{ and } \tau_\gamma = \tau_S \\
(-\tau_T-1)^{-2k} = \omega^{2k} & \text{for } \gamma=T \text{ and } \tau_\gamma = \tau_T \\
(2\tau_U-1)^{-2k} = (-1)^{k} & \text{for } \gamma=U \text{ and } \tau_\gamma = \tau_U\\
(2\tau_{SU}+1)^{-2k} = (-1)^{k} & \text{for } \gamma=SU \text{ and } \tau_\gamma = \tau_{SU}
\end{array} \right.,
\end{eqnarray}
where values of $c$ and $d$ for $\gamma=S,T,U,SU$ are obtained from Eq.~\eqref{eq:STU}. 
Given triplet ($\mathbf{3}$, $\mathbf{3}'$) representation matrices for $S$, $T$ and $U$ in Table~\ref{tab:rep_matrix_main}, it is straightforward to obtain 
\begin{eqnarray} \label{eq:modular_triplet}
&& Y^{(6j+2)}_{\mathbf{3}^{(\prime)}}(\tau_T) \propto \begin{pmatrix} 0 \\ 1 \\ 0 \end{pmatrix}\,,~
Y^{(6j+4))}_{\mathbf{3}^{(\prime)}}(\tau_T) \propto \begin{pmatrix} 0 \\ 0 \\ 1 \end{pmatrix}\,,~
Y^{(6j+6))}_{\mathbf{3}^{(\prime)}}(\tau_T) \propto \begin{pmatrix} 1 \\ 0 \\ 0 \end{pmatrix}\,, \nonumber\\
&& Y^{(2k)}_{\mathbf{3}^{(\prime)}}(\tau_S) \propto \begin{pmatrix} 1 \\ 1 \\ 1 \end{pmatrix} \,, \nonumber\\
&& Y^{(4j+2)}_{\mathbf{3}}(\tau_U) \propto Y^{(4j+4)}_{\mathbf{3}'}(\tau_U)
\propto \begin{pmatrix} 0 \\ 1 \\ -1  \end{pmatrix}\,,\nonumber\\
&& Y^{(4j+2)}_{\mathbf{3}}(\tau_{SU}) \propto Y^{(4j+4)}_{\mathbf{3}'}(\tau_{SU})
\propto \begin{pmatrix} 2 \\ -1 \\ -1  \end{pmatrix}\,,
\end{eqnarray}
where $j$ is a non-negative integer \footnote{Note that for $j=0$, $Y_{\mathbf{3}}^{(2)}$ should be considered since it does not exist. }. These results are obtained without knowing explicit expressions of modular forms. 
However, there are some exceptions of modular forms whose directions cannot be directly determined by the above argument, $Y^{(4j+4)}_{\mathbf{3}}(\tau_U)$, $Y^{(4j+2)}_{\mathbf{3}'}(\tau_U)$ and $Y^{(4j+4)}_{\mathbf{3}}(\tau_{SU})$, $Y^{(4j+2)}_{\mathbf{3}'}(\tau_{SU})$. These modular forms correspond to eigenvectors of degenerate eigenvalues. For instance, $Y^{(2)}_{\mathbf{3}'}(\tau_{SU})$ is the eigenvalue of $\rho_{\mathbf{3}'}(SU)$ with respect to the degenerate eigenvalue $1$. To fully determine the direction of these modular form, we can apply the algebra in Eq.~\eqref{eq:algebra}. Take $Y^{(2)}_{\mathbf{3}'}(\tau_{SU})$ again as an example. It corresponds to the eigenvalue $1$ of $\rho_{\mathbf{3}'}(SU)$. The latter has two linearly independent eigenvectors $(1,1,1)^T$ and $(0,1,-1)^T$, and $Y^{(2)}_{\mathbf{3}'}(\tau_{SU})$ should be a linear combination of them, 
\begin{eqnarray}
Y^{(2)}_{\mathbf{3}'}(\tau_{SU}) = a \begin{pmatrix} 1 \\ 1 \\ 1 \end{pmatrix} + b \begin{pmatrix} 0 \\ 1 \\ -1 \end{pmatrix} \,.
\end{eqnarray}
Taking it to Eq.~\eqref{eq:algebra} \footnote{Although the modular symmetry is broken by the VEV of the modular field. This identity, which is independent of the value of the modular field, is always satisfied.}, we obtain the identity, $[a^2+2(a+b)(a-b)]^2 = [(a+b)^2+2a(a-b)] [(a-b)^2+2a(a+b)]$, which leads to the ratio $b = \pm \sqrt{6} a$. The sign difference, which cannot be determined by the above algebra, is determined by calculating the exact modular functions. Taking the value $\tau_{SU} = -1/2 + i/2$ into the formula of modular forms, we obtain numerically $Y_3 (\tau_{SU}) = -1.09422 i$, $Y_4 (\tau_{SU}) = -3.7745 i$ and $Y_5 (\tau_{SU}) = 1.58606 i$, i.e., $a = -1.09422 i$ and $b = 2.68028 i$. Therefore, we arrive at $b = - \sqrt{6} a$. 
Here, together with $Y^{(2)}_{\mathbf{3}'}(\tau_{SU})$, we list some interesting modular forms respecting to degenerate eigenvalues with modular weights $\leqslant 4$, 
\begin{eqnarray} \label{eq:modular_triplet}
Y^{(2)}_{\mathbf{3}'}(\tau_{SU}) \propto \begin{pmatrix} 1 \\ 1-\sqrt{6} \\ 1+\sqrt{6} \end{pmatrix} \,,~ 
Y^{(4)}_{\mathbf{3}}(\tau_{U}) 
\propto \begin{pmatrix}  \sqrt{2}+ 2 i \\
 \sqrt{2}-i \\
 \sqrt{2}-i 
\end{pmatrix}\,,~
Y^{(4)}_{\mathbf{3}}(\tau_{SU}) 
\propto \begin{pmatrix} 
 \sqrt{2} \\
 \sqrt{2}-\sqrt{3} \\
 \sqrt{2}+\sqrt{3}  \end{pmatrix}\,,
\end{eqnarray}
In addition, we list double modular forms $Y_{\mathbf{2}}^{(2k)}(\tau_S)$, $Y_{\mathbf{2}}^{(2k)}(\tau_U)$ and $Y_{\mathbf{2}}^{(2k)}(\tau_{SU})$, 
\begin{eqnarray} \label{eq:modular_triplet}
Y^{(4j+2)}_{\mathbf{2}}(\tau_U) \propto
Y^{(4j+2)}_{\mathbf{2}}(\tau_{SU}) \propto
\begin{pmatrix} 1 \\ -1  \end{pmatrix} ,\,
Y^{(2k)}_{\mathbf{2}}(\tau_S) \propto
Y^{(4j+4)}_{\mathbf{2}}(\tau_U) \propto 
Y^{(4j+4)}_{\mathbf{2}}(\tau_{SU}) \propto 
\begin{pmatrix} 1 \\ 1  \end{pmatrix} .
\end{eqnarray}

\subsection{Two $S_4$ modular groups}

In our recent paper \cite{deMedeirosVarzielas:2019cyj}, we discussed how to generalise the discussion from a single $S_4$ to multiple $S_4$  modular symmetries. Here we will give a brief review, limiting the discussion to the case of two $S_4$ modular groups relevant to the model discussed later.

Given two infinite modular groups $\overline{\Gamma}^{l}$ and $\overline{\Gamma}^{\nu}$, where the moduli fields are denoted as $\tau_l$ and $\tau_\nu$, respectively. Following Eq.~\eqref{eq:modular_transformation}, any two modular transformations $\gamma_l \times \gamma_\nu$ in $\overline{\Gamma}^l \times \overline{\Gamma}^\nu$ take forms as 
\begin{eqnarray} 
&&\gamma_l \times \gamma_\nu: (\tau_l,\tau_\nu) \to (\gamma_l \tau_l,\gamma_\nu \tau_\nu) = 
\left( \frac{a_l \tau_l + b_l}{c_l \tau_l + d_l}, \frac{a_\nu \tau_\nu + b_\nu}{c_\nu \tau_\nu + d_\nu} \right) \,,
\end{eqnarray}
Two finite modular groups $S_4^l$ and $S_4^\nu$ can be obtained by imposing $T_{\tau_l}^4 = T_{\tau_\nu}^4 = \mathbf{1}$  following the discussion in the former section. Their generators ($S$, $T$, $U$) are denoted by ($S_l$, $T_l$, $U_l$) and ($S_\nu$, $T_\nu$, $U_\nu$), respectively, where the subscripts are only used to distinguish groups. 

The superpotential $W(\phi_i;\tau_l, \tau_\nu)$, which is invariant under any modular transformations, is in general a holomorphic function of the moduli fields $\tau_l$, $\tau_\nu$ and superfields $\phi_i$. 
It is expressed in powers of $\phi_i$ as
\begin{eqnarray}
W(\phi_i;\tau_l, \tau_\nu) = \sum_n  \sum_{\{i_I, \cdots, i_n\}} 
\left(Y_{(I_{Y,l}, I_{Y,\nu})} \phi_{i_1} \cdots \phi_{i_n} \right)_{(\mathbf{1}, \mathbf{1})} \,, 
\end{eqnarray}
the weights of $Y_{(I_{Y,l}, I_{Y,\nu})}$ are given by $k_{Y,l} = k_{1,l}+ \cdots k_{n,l}$ and $k_{Y,\nu} = k_{1,\nu}+ \cdots k_{n,\nu}$. The chiral field $\phi_i$ and
the modular form $Y_{(I_{Y,l}, I_{Y,\nu})}$ respectively transform as 
\begin{eqnarray} \label{eq:field_form_transformation}
 \phi_i(\tau_l, \tau_\nu) &\to& \phi_i(\gamma_l\tau_l, \gamma_\nu \tau_\nu) 
\nonumber\\
&& = (c_l\tau_l + d_l)^{-2k_{i,l}}  (c_\nu\tau_\nu + d_\nu)^{-2k_{i,\nu}} \rho_{I_{i,l}}(\gamma_l) \phi_i(\tau_l, \tau_\nu) \rho^T_{I_{i,\nu}}(\gamma_\nu)\,, \nonumber\\
\hspace{-5mm}
Y_{(I_{Y,l}, I_{Y,\nu})}(\tau_l, \tau_\nu) &\to& Y_{(I_{Y,l}, I_{Y,\nu})}(\gamma_l \tau_l, \gamma_\nu \tau_\nu) \nonumber\\
&&= (c_l\tau_l + d_l)^{2k_{Y,l}} (c_\nu \tau_\nu + d_\nu)^{2k_{Y,\nu}}
 \rho_{I_{Y,l}}(\gamma_l) Y_{(I_{Y,l}, I_{Y,\nu})}(\tau_l, \tau_\nu) \rho^T_{I_{Y,\nu}}(\gamma_\nu) \,.
\end{eqnarray}
Here, we have arranged $\phi_i$ and $Y_{(I_{Y,l}, I_{Y,\nu})}$ as matrices, and let $\gamma_l$ act on them vertically and $\gamma_\nu$ act on them horizontally. 

Including two modular symmetries allows us to break modular symmetries into different subgroups in charged lepton sector and neutrino sector respectively. For example, $\langle \tau_l \rangle = \tau_T$ and $\langle \tau_\nu \rangle = \tau_S$ or $\langle \tau_\nu \rangle = \tau_U$. We will discuss phenomenological consequences of these different breaking chains in the next section in the model building. 

\section{A minimal model with $S_4^l \times S_4^\nu$ modular symmetries}
\label{model}

\begin{table}[h] 
\begin{center}
{\begin{tabular}{| l | c c c c|}
\hline \hline
Fields & $S_4^l$ & $S_4^\nu$ & $2k_l$ & $2k_\nu$\\ 
\hline \hline
$e^c$ & $\mathbf{1}'$ & $\mathbf{1}$ & $-6$ & $-2$ \\
$\mu^c$ & $\mathbf{1}'$ & $\mathbf{1}$ & $-4$ & $-2$ \\
$\tau^c$ & $\mathbf{1}'$ & $\mathbf{1}$ & $-2$ & $-2$ \\
$L$ & $\mathbf{3}$ & $\mathbf{1}$ & 0 & $+2$\\
$\nu^c$ & $\mathbf{1}$ & $\mathbf{3}$ & 0 & $-2$ \\
\hline 
$\Phi$ & $\mathbf{3}$ & $\mathbf{3}$ & 0 & $0$ \\
\hline 
$H_{u,d}$ & $\mathbf{1}$ & $\mathbf{1}$ & 0 & $0$ \\
\hline \hline
\end{tabular}
\hspace{2mm}
\begin{tabular}{| l | c c c c|}
\hline \hline
Yukawas / masses &$S_4^l$ & $S_4^\nu$ & $2k_l$ & $2k_\nu$\\
\hline \hline
$Y_e(\tau_l)$ & $\mathbf{3}'$ & $\mathbf{1}$ & $+6$ & 0 \\
$Y_\mu(\tau_l)$ & $\mathbf{3}'$ & $\mathbf{1}$ & $+4$ & 0 \\
$Y_\tau(\tau_l)$ & $\mathbf{3}'$ & $\mathbf{1}$ & $+2$ & 0 \\\hline
$M_{\mathbf{1}}(\tau_\nu)$ & $\mathbf{1}$ & $\mathbf{1}$ & 0 & $+4$ \\
$M_{\mathbf{2}}(\tau_\nu)$ & $\mathbf{1}$ & $\mathbf{2}$ & 0 & $+4$ \\
$M_{\mathbf{3}}(\tau_\nu)$ & $\mathbf{1}$ & $\mathbf{3}$ & 0 & $+4$ \\
\hline \hline 
\multicolumn{5}{c}{}
\end{tabular}}
\end{center}
\caption{Transformation properties of leptons, Yukawa couplings and right-handed neutrino masses in $S_4^l \times S_4^\nu$. \label{tab:particle_contents}}
\end{table}

The extension from one single modulus field to multiple moduli fields \cite{deMedeirosVarzielas:2019cyj}
opens the door to new directions in modular model building. 
Following the approach of multiple modular symmetries, we will construct a flavour model with two modular symmetries, $S_4^l$ and $S_4^\nu$, with moduli fields labelled by $\tau_l$ and $\tau_\nu$, respectively. After moduli fields gain different VEVs, different textures of mass matrices are realised in charged lepton and neutrino sectors.

The transformation properties of leptons are given in Table~\ref{tab:particle_contents}. Leptons, including right-handed neutrinos $\nu^c$ are arranged in the following way: 1) the right-handed leptons $e^c$, $\mu^c$ and $\tau^c$ are singlets $\mathbf{1}'$ of $S_4^l$ and trivial singlets $\mathbf{1}$ of $S_4^\nu$, and have different weights $2k_l=-6,-4,-2$, respectively and the same weight $2k_\nu=-2$; 2) the lepton doublets $L$ form a triplet of $S_4^l$ with zero weight, but a singlet of $S_4^\nu$ with weight $2k_\nu=+2$; 3) We introduce three right-handed neutrinos $\nu^c$ which form a triplet of $S_4^\nu$ with weight $2k_\nu=-2$.

Superpotential terms for generating charged lepton and neutrino mass matrices are respectively given by
\begin{eqnarray} \label{eq:superpotential}
w &=& \left[ L Y_e(\tau_l) e^c + L Y_\mu(\tau_l) \mu^c + L Y_\tau(\tau_l) \tau^c \right] H_d \nonumber\\
&+& \frac{y_\nu}{\Lambda} L \Phi  \nu^c H_u+ \frac{1}{2} M_{\mathbf{1}}(\tau_\nu) (\nu^c \nu^c)_{\mathbf{1}} + \frac{1}{2} M_{\mathbf{2}}(\tau_\nu) (\nu^c \nu^c)_{\mathbf{2}} + \frac{1}{2} M_{\mathbf{3}}(\tau_\nu) (\nu^c \nu^c)_{\mathbf{3}}\,. 
\end{eqnarray}
To be invariant under the modular transformation, 
$Y_{e,\mu,\tau}$ are $\mathbf{3}'$-plet modular forms of $S_4^l$ with weights $2k_l=6,4,2$, respectively, $y_\nu$ can only be a modulus-independent coefficient in this model instead of a modular form. Masses for right-handed neutrinos all takes the same modular weight $2k_\nu=+4$. $M_{\mathbf{1}}(\tau_\nu)$, $M_{\mathbf{2}}(\tau_\nu)$ and $M_{\mathbf{3}}(\tau_\nu)$ represents $\mathbf{1}$-, $\mathbf{2}$- and $\mathbf{3}$-plets modular forms appearing in right-handed neutrino mass terms. The dimension-five operator $L \Phi  \nu^c H_u$ is understood as an effective operator after integrating out heavy particles. A typical example is including a pair of electroweak-neutral superfields, $F, F^c$, with couplings $L F^c H_u+M_F F F^c + F \Phi \nu^c$, where $F, F^c \sim (\mathbf{3}, \mathbf{1})$ of $(S_4^l, S_4^\nu)$ and $(2 k_l, 2k_\nu) = (0, \pm 2)$. Decoupling of these fields introduces no additional relevant dimension-five operator but the one in Eq.~\eqref{eq:superpotential}.

\subsection{$S_4^l \times S_4^\nu \to S_4$}

In order to achieve this breaking we have introduced a scalar $\Phi$, which is arranged as a bi-triplet, i.e., $\Phi \sim (\mathbf{3}, \mathbf{3})$ of $S_4^l \times S_4^\nu$, and its modular weights $2k_l$ and $2k_\nu$ are arranged at zero. This scalar is not supposed to generate special Yukawa textures for leptons. Instead, it is used for the connection between two $S_4$'s and its VEV is the key to break two $S_4$'s to a single $S_4$. This idea and relevant technique for how to obtain the required the VEV was introduced and developed in \cite{deMedeirosVarzielas:2019cyj}. We will not repeat them in this article. Without loss of generality, we can fix the VEV of $\Phi$ at $\langle \Phi \rangle_{\alpha i} = v_\Phi (P_{23})_{\alpha i}$ with
\begin{eqnarray}
P_{23} = \begin{pmatrix} 1 & 0 & 0 \\ 0 & 0 & 1 \\ 0 & 1 & 0 \end{pmatrix} \,.
\end{eqnarray} 
Here, $\alpha=1,2,3$ corresponds the entries of the triplet of $S_4^l$, while $i=1,2,3$ corresponds to those of $S_4^\nu$.
With this VEV, we can realise the breaking 
$S_4^l \times S_4^\nu \to S_4$.

As mentioned, $S_4^l \times S_4^\nu$ is broken after $\Phi$ gains the above VEV. 
The scalar $\Phi$ connects $S_4^l$ with $S_4^\nu$ via the effective dimension-5 operator $\frac{y_\nu}{\Lambda} L \Phi  \nu^cH_u$, 
responsible for Dirac neutrino
Yukawa couplings. This operator is explicitly expanded as 
\begin{eqnarray}
\frac{y_\nu}{\Lambda} (L_1, L_2, L_3) P_{23} \begin{pmatrix}
\Phi_{11} & \Phi_{12} & \Phi_{13} \\ 
\Phi_{21} & \Phi_{22} & \Phi_{23} \\ 
\Phi_{31} & \Phi_{32} & \Phi_{33} 
\end{pmatrix} P_{23} 
\begin{pmatrix}
\nu^c_{1} \\ 
\nu^c_{2} \\ 
\nu^c_{3} 
\end{pmatrix} H_u\,.
\end{eqnarray}
Given the VEV $\langle \Phi \rangle = P_{23} v_{\Phi}$, this term is not invariant under transformations $\gamma_l$ and $\gamma_\nu$ of $S_4^l$ and $S_4^\nu$ and thus the modular symmetry $S_4^l \times S_4^\nu$ is broken. 
However, given any $\gamma_l$ of $S_4^l$, we can perform the same transformation $\gamma_\nu = \gamma_l$ of $S_4^\nu$, such that the VEV of $\Phi$ keeps invariant, namely,
\begin{eqnarray} 
\langle \Phi \rangle \to \rho_{\mathbf{3}}(\gamma_l)\langle \Phi \rangle \rho^T_{\mathbf{3}}(\gamma_\nu) = \langle \Phi \rangle
\end{eqnarray}
for $\gamma_l = \gamma_\nu$. This equation is simply proven after we write it in the following matrix form
\begin{eqnarray} 
\rho_{\mathbf{3}}(\gamma_l) 
\langle \Phi \rangle
\rho^T_{\mathbf{3}}(\gamma_\nu) = 
\rho_{\mathbf{3}}(\gamma_l) P_{23}
\rho^T_{\mathbf{3}}(\gamma_\nu) v_\Phi 
= \rho_{\mathbf{3}}(\gamma_l \gamma_\nu^{-1}) P_{23}
 v_\Phi  = \rho_{\mathbf{3}}(\gamma_l \gamma_\nu^{-1}) \langle \Phi \rangle  \,,
\end{eqnarray}
where $P_{23} \rho^T_{\mathbf{3}} (\gamma) =\rho_{\mathbf{3}} (\gamma^{-1}) P_{23}$ has been used. It is obvious that $\langle \Phi \rangle$ is invariant if $\gamma_l = \gamma_\nu$. Therefore, the diagonal part of $S_4^l\times S_4^\nu$ is preserved in the vacuum. 
$\frac{y_\nu}{\Lambda} L \Phi  \nu^cH_u$ is the only term which breaks $S_4^l \times S_4^\nu$ to a single $S_4$. Fix $\Phi$ at its VEV, this term is left with $y_D (L_1 \nu^c_1 + L_2 \nu^c_3 + L_3 \nu^c_2) H_u$, where we have denoted $y_D = y_\nu v_\Phi / \Lambda$. It appears as a renormalisable Dirac neutrino Yukawa interaction at low energy, which is proportional to $P_{23}$. Therefore all neutrino mixing arises from the heavy Majorana neutrino mass matrix.

To summarise, after $\Phi$ gains the VEV, superpotential $w$ is effectively given by
\begin{eqnarray}
w_{\rm eff} &=& \left[ L Y_e(\tau_l) e^c + L Y_\mu(\tau_l) \mu^c + L Y_\tau(\tau_l) \tau^c \right] H_d \nonumber\\
&+& y_D L \nu^c H_u+ \frac{1}{2} M_{\mathbf{1}}(\tau_\nu) (\nu^c \nu^c)_{\mathbf{1}} + \frac{1}{2} M_{\mathbf{2}}(\tau_\nu) (\nu^c \nu^c)_{\mathbf{2}} + \frac{1}{2} M_{\mathbf{3}}(\tau_\nu) (\nu^c \nu^c)_{\mathbf{3}}\,. 
\end{eqnarray}
The full effective superpotential involves two moduli fields. It is not invariant in $S_4^l \times S_4^\nu$ but their diagonal subgroup $S_4$.

Under this symmetry, a modular transformation appears to be
\begin{eqnarray} \label{eq:modular_transformation_D}
\gamma: (\tau_l,\, \tau_\nu) \to (\gamma\tau_l, \gamma\tau_\nu) = \left( \frac{a \tau_l + b}{c \tau_l + d}, \frac{a \tau_\nu + b}{c \tau_\nu + d} \right)
\end{eqnarray}
for any $\gamma \in S_4$. 
We also write out transformation properties of leptons 
\begin{eqnarray}
 L(\tau_\nu) &\to& L(\gamma\tau_\nu) = (c \tau_\nu + d)^{2} \rho_{\mathbf{3}}(\gamma) L(\tau_\nu) \,,\nonumber\\
 \alpha^c(\tau_l,\tau_\nu) &\to& \alpha^c(\gamma\tau_l,\gamma\tau_\nu) = (c \tau_l + d)^{-2k_\alpha} (c \tau_\nu + d)^{-2} \alpha^c(\tau_l,\tau_\nu) \,,\nonumber\\
 \nu^c(\tau_\nu) &\to& \nu^c(\gamma\tau_\nu) = (c \tau_\nu + d)^{-2} \rho_{\mathbf{3}}(\gamma) \nu^c (\tau_\nu) \,,
  \label{eq:field_transformation_D}
\end{eqnarray}
and those for modular forms
\begin{eqnarray}
 Y_\alpha(\tau_l) &\to& Y_\alpha(\gamma\tau_l) = (c \tau_l + d)^{2k_\alpha} \rho_{\mathbf{3}}(\gamma) Y_\alpha(\tau_l) \,,\nonumber\\
M_{\mathbf{r}}(\tau_\nu) &\to& M_{\mathbf{r}}(\gamma \tau_\nu) = (c \tau_\nu + d)^4 \rho_{\mathbf{r}}(\gamma)  M_{\mathbf{r}}(\tau_\nu) \,, \label{eq:form_transformation_D}
\end{eqnarray}
where $\alpha=e,\mu,\tau$, $k_{e,\mu,\tau} = 3,2,1$ and $\mathbf{r}=\mathbf{1}, \mathbf{2}, \mathbf{3}$. Note that in the residual $S_4$ symmetry, we have not induced any correlation between the moduli fields $\tau_l$ and $\tau_\nu$. Namely, $\tau_l$ and $\tau_\nu$ can gain independent VEVs. 
Furthermore, there is no flavon fields involved in the effective superpotential. 

\begin{figure}[ht]
\centering

\hspace*{1ex}
	\includegraphics[width=0.4\textwidth]{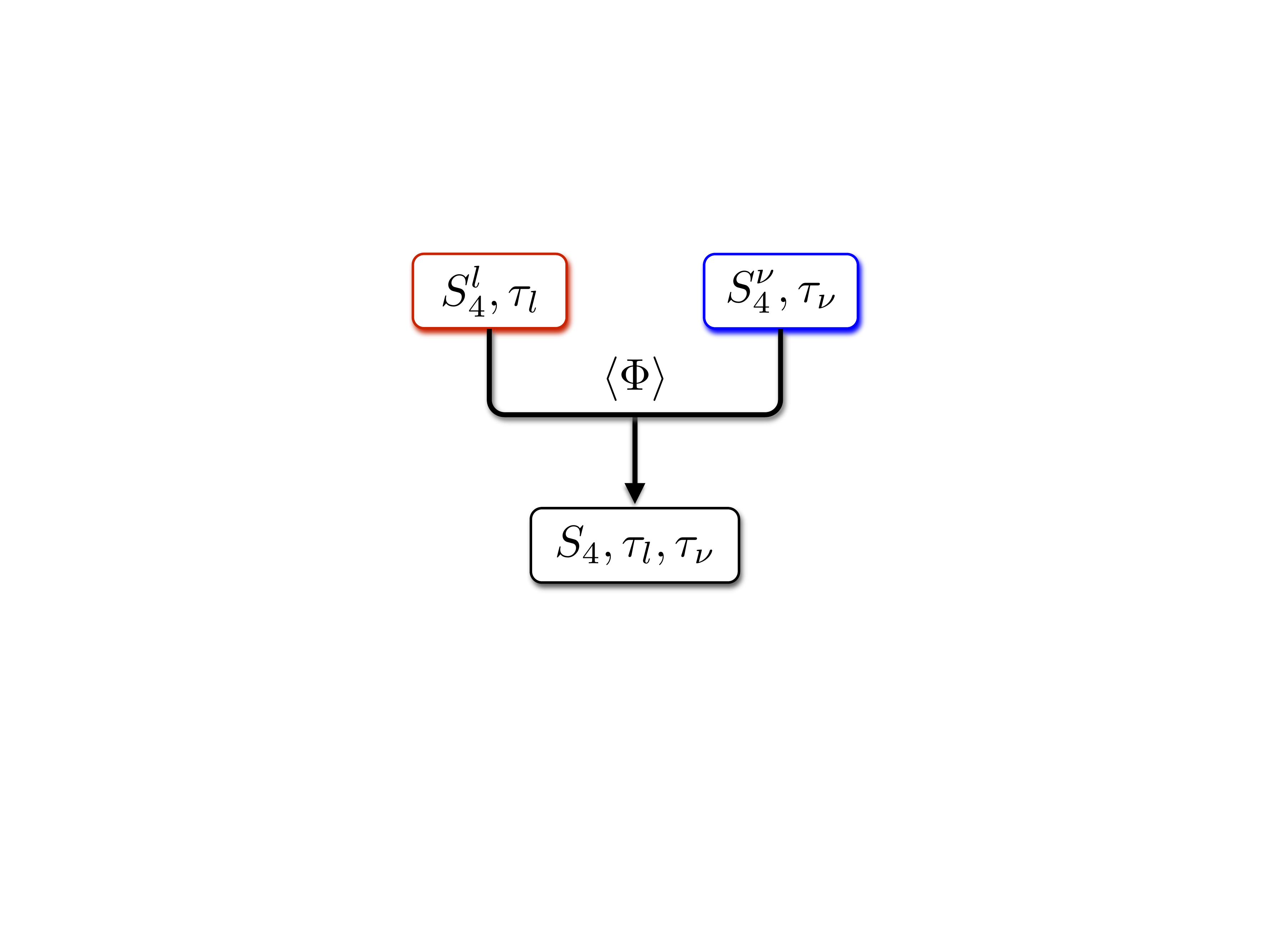}
\caption{Diagram of the breaking of $S_4^l \times S_4^\nu \to S_4$, their diagonal subgroup,
through the VEV of $\Phi$.}
\label{fig:S4s}
\end{figure}

Geometically, we represent the idea of $S_4^l \times S_4^\nu \to S_4$ in the sketch shown in Fig. \ref{fig:S4s}. 

\subsection{Flavour structure after $S_4$ breaking}

In the charged lepton sector, we assume the VEV of $\tau_l$ fixed at $\langle \tau_l \rangle = \tau_T =\omega$, which is a stabiliser of $T$. At this stabiliser, a residual modular $Z_3^T$ symmetry is preserved in the charged lepton sector. It has been proven in \cite{deMedeirosVarzielas:2019cyj} that $Y_{e,\mu,\tau}(\tau_T)$ are eigenvectors of the $3 \times 3$ representation matrix of $T$ for eigenvalues $1$, $\omega$ and $\omega^2$, respectively. Namely, the Yukawa coupling vectors are 
\begin{eqnarray}
Y_e(\tau_T) \propto
\begin{pmatrix}
1\\
0\\
0
\end{pmatrix} \,,~
Y_\mu(\tau_T) \propto
\begin{pmatrix}
0\\
0\\
1
\end{pmatrix}\,,~
Y_\tau(\tau_T) \propto
\begin{pmatrix}
0\\
1\\
0
\end{pmatrix}\,,
\end{eqnarray}
for weights $2k_l=6, 4, 2$, respectively. These modular forms will lead to diagonal Yukawa couplings for the charged leptons. 
We have also seen that the Dirac neutrino Yukawa matrix is proportional to $P_{23}$.
Therefore all lepton mixing arises from the heavy Majorana neutrino mass matrix, to which we now turn.

In the neutrino sector, the right-handed neutrino mass matrix is explicitly written to be
\begin{eqnarray} \label{eq:superpotential_2}
M_R = 
\begin{pmatrix}
M_{\mathbf{1}} & 0 & 0 \\ 
0 & 0 & M_{\mathbf{1}} \\ 
0 & M_{\mathbf{1}} & 0
\end{pmatrix} + 
\begin{pmatrix}
0 & M_{\mathbf{2},1} & M_{\mathbf{2},2} \\ 
M_{\mathbf{2},1} & M_{\mathbf{2},2} & 0 \\ 
M_{\mathbf{2},2} & 0 & M_{\mathbf{2},1}
\end{pmatrix}
+
\begin{pmatrix}
2 M_{\mathbf{3},1} & -M_{\mathbf{3},3} & -M_{\mathbf{3},2} \\ -M_{\mathbf{3},3} & 2M_{\mathbf{3},2} & -M_{\mathbf{3},1} \\ -M_{\mathbf{3},2} & -M_{\mathbf{3},1} & 2M_{\mathbf{3},3}
\end{pmatrix}
\,, 
\end{eqnarray}
where $M_{\mathbf{r},i}$ is the $i$-th component of $M_{\mathbf{r}}(\tau)$, $i=1,2$ for $\mathbf{r}=\mathbf{2}$ and $i=1,2,3$ for $\mathbf{r}=\mathbf{3}$. The Dirac mass matrix is trivially given by 
\begin{eqnarray} \label{eq:superpotential_2}
M_D = 
y_D P_{23} v_u\,,
\end{eqnarray}
The active neutrino mass matrix is obtained by applying the seesaw formula 
\begin{eqnarray}
M_\nu = - M_D M_R^{-1} M_D^T = -y_D^2 v_u^2 P_{23} M_R^{-1} P_{23}\,.
\label{seesaw}
\end{eqnarray}
Specifically, the mass eigenvalues of $M_\nu$, $m_i$ for $i=1,2,3$, are given by $m_i = y_D^2 v_u^2 /M_i$. The (1,1) entry of $M_\nu$ gives rise to the effective mass parameter in neutrinoless double beta decay $m_{ee} \equiv |(M_\nu)_{(1,1)}| = y_D^2 v_u^2 |(M_R^{-1})_{(1,1)}|$.

Since the charged lepton mass matrix is diagonal, the PMNS matrix is determined by the structure of neutrino mass matrix which is governed by the VEV of $\tau_\nu$. 
We assume the stabiliser in the neutrino sector 
\footnote{Note that if we had selected $\langle \tau_\nu \rangle = \tau_{S} = i\infty$, we would have obtained $M_{\mathbf{2}}(\tau_S) \propto (1,1)^T$ and $M_{\mathbf{3}}(\tau_S) \propto (1,1,1)^T$, with a residual {\it $Z_2^S$ flavour symmetry} preserved in the neutrino sector \cite{deMedeirosVarzielas:2019cyj}, leading to tri-bimaximal mixing.
Alternatively the choice $\langle \tau_\nu \rangle$ by $\tau_{U} = 1/2+i/2$ would preserve a residual $Z_2^U$ flavour symmetry
corresponding to a mu-tau permutation symmetry in the neutrino sector.
Since both patterns are excluded, due to the prediction of vanishing $\theta_{13}$, we will not discuss them any further here. }
, $\langle \tau_\nu \rangle = \tau_{SU} = -\frac{1}{2}+\frac{i}{2}$. At this stabiliser, we are left with a {\it residual $Z_2^{SU}$ symmetry}. In former discussion in the framework of flavour symmetry,  the $Z_2^{SU}$ residual symmetry is crucial to realise the TM$_1$ mixing \cite{deMedeirosVarzielas:2019cyj}. 
$M_{\mathbf{2}}$ and $M_{\mathbf{3}}$ take directions $M_{\mathbf{2}} \propto (1,1)^T$ and $M_{\mathbf{3}} \propto (\sqrt{2},\sqrt{2}-\sqrt{3} ,\sqrt{2}+\sqrt{3})^T$, respectively. Together with $M_{\mathbf{1}}$, we write them in the following way, 
\begin{eqnarray}
M_{\mathbf{1}}(\tau_{SU}) = a\,,~
M_{\mathbf{2}}(\tau_{SU}) = b
\begin{pmatrix}
1\\
1
\end{pmatrix} \,,~
M_{\mathbf{3}}(\tau_{SU}) = c
\begin{pmatrix}
\sqrt{2}\\
\sqrt{2} - \sqrt{3} \\
\sqrt{2} + \sqrt{3}
\end{pmatrix}\,.
\end{eqnarray}
Thus, the Majorana mass matrix for right-handed neutrinos are written in the form
\begin{eqnarray} 
M_R = a \begin{pmatrix}
1 & 0 & 0 \\ 0 & 0 & 1 \\ 0 & 1 & 0
\end{pmatrix} + 
b \begin{pmatrix}
0 & 1 & 1 \\ 1 & 1 & 0 \\ 1 & 0 & 1
\end{pmatrix}
+ c \sqrt{2}
\begin{pmatrix}
2 & -1 & -1 \\ -1 & 2 & -1 \\ -1 & -1 & 2
\end{pmatrix}
-c \sqrt{3}
\begin{pmatrix}
0 & 1 & -1 \\ 
1 & 2 & 0 \\ -1 & 0 & -2
\end{pmatrix} \,,
\label{MR}
\end{eqnarray}
where $a$, $b$ and $c$ are complex parameters. 
As discussed in the next subsection, the above heavy Majorana neutrino mass matrix, together with a Dirac neutrino Yukawa matrix proportional to $P_{23}$,
and a diagonal charged lepton mass matrix, will lead to Trimaximal TM$_1$ lepton mixing which preserves the first column on the tri-bimaximal
mixing matrix,
\begin{eqnarray}
U_{\rm TM_1}=\left(
\begin{array}{ccc}
 \frac{2}{\sqrt{6}} & - & - \\
 -\frac{1}{\sqrt{6}} &- &- \\
 -\frac{1}{\sqrt{6}} & - & - \\
\end{array}
\right)\,.
\label{TM1}
\end{eqnarray}
It is worth mentioning that in classical flavour models without modular symmetry, such as  \cite{Luhn:2013vna},
coefficients for the third and fourth terms on the right hand side of Eq.~\eqref{MR} are fully arbitrary, but here they are constrained by a fixed ratio $-\sqrt{2/3}$. 
Thus, in the modular symmetry model here, $M_R$ depends on three complex parameters, while in the classical (non-modular symmetry) model in \cite{Luhn:2013vna}
$M_R$ depends on four complex parameters. We will show that having fewer parameters leads to a new neutrino mass sum rule, not present in the previous flavon models
of TM$_1$ mixing which do not rely on modular symmetry.

\subsection{Results for neutrino mass and mixing}

The heavy Majorana mass matrix $M_R$ in Eq.~\eqref{MR} can be put into block diagonal form by applying the TBM mixing matrix,
\begin{eqnarray} \label{eq:MR_TBM}
U_{\rm TBM}^T M_R U_{\rm TBM} = 
\left(
\begin{array}{ccc}
 -\beta-2\gamma & 0 & 0 \\
 0 & \alpha & \gamma \\
 0 & \gamma & \beta \\
\end{array}
\right) \,,
\end{eqnarray}
where $\alpha=a+2b$, $\beta=b-a+3\sqrt{2}c$ and $\gamma=-3\sqrt{2}c$. 
Since the remaining (2,3) rotations required to diagonalise $M_R$ leave the first column of 
the TBM matrix unchanged, this implies that $M_R$ is diagonalised by the TM$_1$ matrix in Eq.~\eqref{TM1}.
Then, since the Dirac neutrino Yukawa matrix proportional to $P_{23}$, the seesaw mass matrix $M_\nu$ in Eq.~\eqref{seesaw}
will also be diagonalised by $U_{\rm TM_1}$. Hence, as claimed, we have trimaximal TM$_1$ lepton mixing, given that the charged lepton mass matrix is diagonal.

Returning to Eq.~\eqref{eq:MR_TBM},
the re-parametrised mass parameters $\alpha$, $\beta$ and $\gamma$ are independent complex parameters. Namely, the bottom right $2 \times 2$ submatrix  in Eq.~\eqref{eq:MR_TBM} is an arbitrary complex symmetric matrix. Thus it can be diagonalised by a $2 \times 2$  unitary matrix
\begin{eqnarray} 
V = e^{i \alpha_3}
\begin{pmatrix} \cos\theta_R e^{- i\alpha_1} & \sin\theta_R e^{i\alpha_2} \\  \sin\theta_R e^{-i\alpha_2} & -\cos\theta_R e^{i\alpha_1} \end{pmatrix} 
\end{eqnarray}
with two real eigenvalues $M_2$ and $M_3$. Here, $M_2$, $M_3$ and $V$ are arbitrary. However, the first eigenvalue of $M_R$, i.e., $M_1$, is not arbitrary, but determined by $M_2$, $M_3$ and the mixing parameters in $V$ via
\begin{eqnarray} \label{eq:M1}
M_1 = |\beta+2\gamma| =  \left| M_2 (\sin^2\theta_R e^{i2\alpha_2}
+ \sin 2 \theta_R e^{i (\alpha _1+\alpha _2)})
+M_3 (\cos^2\theta_R e^{-i2\alpha_1} - e^{-i(\alpha _1+\alpha _2)}) \right|\,.
\end{eqnarray}

According to the above discussion, the model predicts lepton mixing to be of the TM$_1$ form, $U_{\rm PMNS}=U_{\rm TM_1}$, with 
the general form of TM$_1$ mixing in Eq.~\eqref{TM1} parametrised as 
\begin{eqnarray}
U_{\rm TM_1} = U_{\rm TBM} \begin{pmatrix} e^{\alpha_3'} & 0 & 0 \\ 0 & \cos\theta_R e^{i\alpha_1} & \sin\theta_R e^{- i\alpha_2} \\ 0 & -\sin\theta_R e^{i\alpha_2} & \cos\theta_R e^{- i\alpha_1} \end{pmatrix} \,,
\end{eqnarray}
where $\alpha'_3 = \frac{1}{2} {\rm arg}(-\beta - 2\gamma)$. The mixing angles and Dirac-type CP-violating phase are determined to be \cite{Luhn:2013vna}
\begin{eqnarray}
\sin\theta_{13} &=& \frac{\sin \theta_R}{\sqrt{3}} \,, \nonumber\\
\tan\theta_{12} &=& \frac{\cos \theta_R}{\sqrt{2}} \,, \nonumber\\
\tan\theta_{23} &=& \left|\frac{ \cos \theta_R + \sqrt{\frac{2}{3}} e^{i (\alpha_1-\alpha_2)} \sin \theta_R}{ \cos \theta_R- \sqrt{\frac{2}{3}} e^{i (\alpha_1-\alpha_2)} \sin \theta_R} \right| \,, \nonumber\\
\delta &=& {\rm arg}\left[ (5 \cos 2 \theta_R +1) \cos ( \alpha_1 - \alpha_2 )-i (\cos 2 \theta_R + 5) \sin ( \alpha_1 - \alpha_2 ) \right]
 \,.
\end{eqnarray}

The above $\rm{TM}_1$ mixing implies three equivalent relations:
\be
\tan \theta_{12} = \frac{1}{\sqrt{2}}\sqrt{1-3s^2_{13}}\ \ \ \ {\rm or} \ \ \ \ 
\sin \theta_{12}= \frac{1}{\sqrt{3}}\frac{\sqrt{1-3s^2_{13}}}{c_{13}} \ \ \ \ {\rm or} \ \ \ \ 
\cos \theta_{12}= \sqrt{\frac{2}{3}}\frac{1}{c_{13}}
\label{t12p}
\ee
leading to a prediction $\theta_{12}\approx 34^{\circ}$,
in excellent agreement with current global fits, assuming $\theta_{13}\approx 8.5^{\circ}$.
By contrast, the corresponding $\rm{TM}_2$ relations imply $\theta_{12}\approx 36^{\circ}$ \cite{Albright:2008rp}, which is on the edge of the three sigma region, and hence disfavoured by current data.
$\rm{TM}_1$ mixing also leads to an exact sum rule relation relation for $\cos \delta$ in terms of the other lepton mixing angles
\cite{Albright:2008rp},
\be
\cos \delta = - \frac{\cot 2\theta_{23}(1-5s^2_{13})}{2\sqrt{2}s_{13}\sqrt{1-3s^2_{13}}},
\label{TM1sum}
\ee
which, for approximately maximal atmospheric mixing, predicts $\cos \delta \approx 0$,
$\delta \approx \pm 90^{\circ}$.
Such {\em atmospheric mixing sum rules} may be tested
in future experiments~\cite{Ballett:2013wya}.

Apart from predicting TM$_1$ lepton mixing, the
model also predicts a neutrino mass sum rule \cite{King:2013psa} between the light physical effective Majorana neutrino 
mass eigenvalues $m_i$ (i.e. the active neutrino masses relevant for low energy experiments).
Using the correlation of $M_1$ and $M_{2,3}$ in Eq.~\eqref{eq:M1} and $M_i = - y_D^2 v_u^2 /m_i$ for $i=1,2,3$, we obtain a new neutrino mass sum rule for 
the active neutrino masses (beyond those reported in \cite{King:2013psa}),
\begin{eqnarray} 
\frac{1}{m_1} &=& \left| \frac{1}{m_2} \left(\sin^2\theta_R e^{-i2\alpha_2} + \sin 2 \theta_R e^{-i (\alpha _1+\alpha _2)} \right) + 
\frac{1}{m_3} \left(\cos^2\theta_R e^{i2\alpha_1} - \sin 2 \theta_R e^{i(\alpha _1+\alpha _2)} \right) \right|\,. \nonumber\\
\end{eqnarray}
Furthermore, we can predict the effective neutrino mass parameter $m_{ee}$ in neutrino-less double beta decay experiments. It is effectively represented as 
\begin{eqnarray}
m_{ee} &=& y_D^2 v_u^2 |(M_R^{-1})_{(1,1)}| 
=   y_D^2 v_u^2 \left| \frac{2}{3 (\beta +2 \gamma )}-\frac{\beta }{3 (\alpha  \beta - \gamma^2)} \right| \nonumber\\
&=& \Big| \frac{2 m_2 m_3}{ 3 \left(
m_2 (\cos^2\theta_R e^{i2\alpha_1} - \sin 2 \theta_R e^{i(\alpha _1+\alpha _2)} )
+ m_3 ( \sin^2\theta_R e^{-i2\alpha_2} + \sin 2 \theta_R e^{-i (\alpha _1+\alpha _2)} ) \right) 
} \Big.
 \nonumber\\
&&+ \Big. \frac{1}{3} ( m_2 \cos ^2 \theta_R e^{2 i \alpha_1} + m_3 \sin^2 \theta_R e^{-2 i \alpha_2} ) \Big| \,,
\end{eqnarray}

\begin{figure}[ht]
\begin{minipage}[t]{0.5\textwidth}
\centering
	\includegraphics[width=1\textwidth]{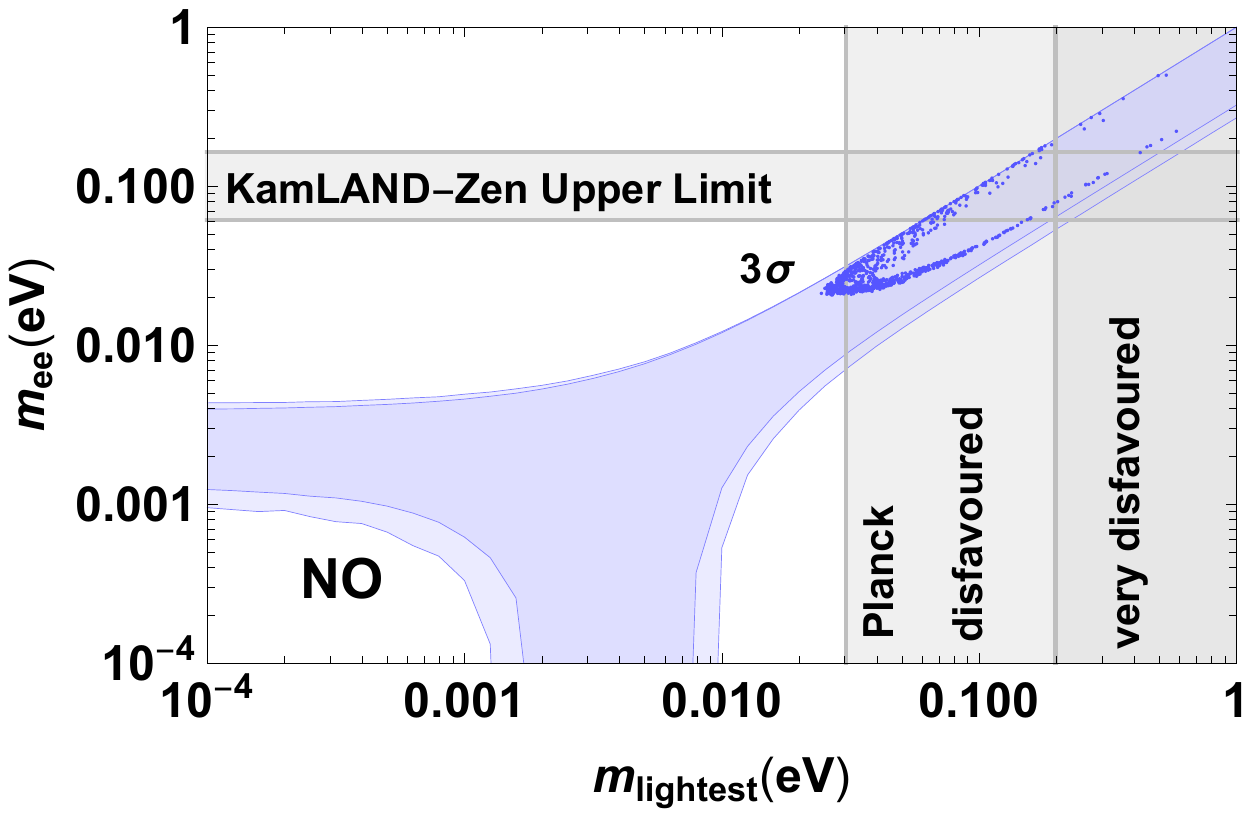}
\end{minipage}
\begin{minipage}[t]{0.5\textwidth}
        \includegraphics[width=1\textwidth]{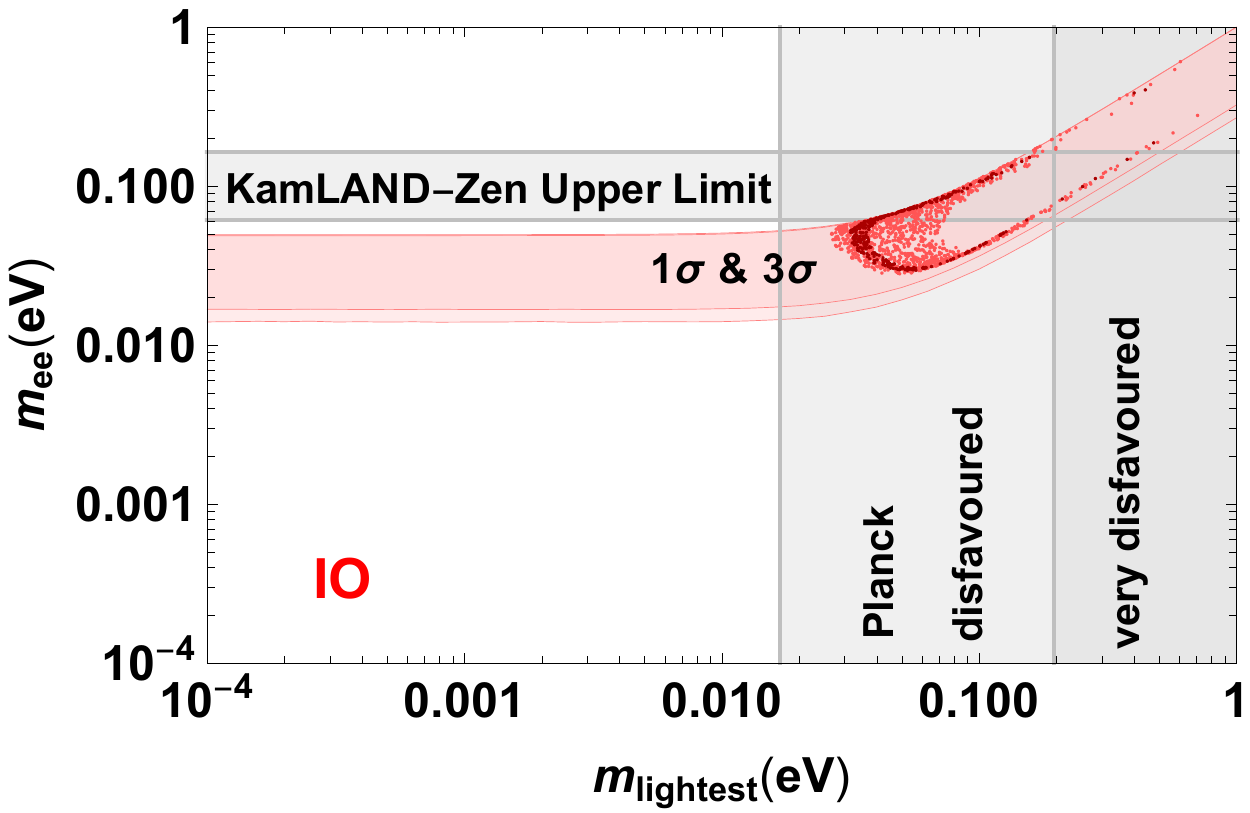}
\end{minipage}
\caption{Predictions of $m_{\text{lightest}}$ vs $m_{ee}$ for both normal ordering (NO, left panel) and inverted ordering (IO, right panel) of neutrino masses, 
allowed by the model, where $m_{\text{lightest}} = m_1$ for NO and $m_{\text{lightest}} = m_3$ for IO. $1\sigma$ and $3\sigma$ range data of oscillation parameters from \cite{Esteban:2018azc, nufit4} are taken as inputs. The general parameter space of $m_{ee}$ allowed by oscillation data and current upper limit from KamLAND-Zen and cosmological constraints from PLANCK 2018 \cite{Aghanim:2018eyx} (disfavoured region $0.12\,{\rm eV} < \sum m_i < 0.60\,{\rm eV}$ and very disfavoured region $\sum m_i > 0.60\,{\rm eV}$) are shown for comparison. }
\label{fig:mee}
\end{figure}

In Fig.~\ref{fig:mee} we display the prediction of $m_{\text{lightest}}$ vs $m_{ee}$, where $m_{\text{lightest}} = m_1$ for neutrino masses with normal ordering (NO) and $m_{\text{lightest}} = m_3$ for inverted ordering (IO). $1\sigma$ and $3\sigma$ ranges of oscillation parameters from \cite{Esteban:2018azc, nufit4} have been taken as inputs in the left and right panels, respectively.
In this plot, we also show the upper limit from KamLAND-Zen experiment, $(m_{ee})_{\text{upper limit}}=0.061 - 0.165$ eV, which is the current best experimental constraint for $m_{ee}$, and cosmological constraints from Planck 2018 \cite{Aghanim:2018eyx}, for comparison. The latter set limits on $\sum_i m_i$. Depending on data inputs, different limits are obtained. In the figure, we consider two limits $\sum_i m_i<0.12\,{\rm eV}$ (95\%, Planck TT,TE,EE+lowE+lensing+BAO+$\theta_{\rm MC}$) and $\sum_i m_i<0.60\,{\rm eV}$ (95\%, Planck lensing+BAO+$\theta_{\rm MC}$), which we refer to ``disfavoured'' and ``very disfavoured'' regimes, respectively. The first limit was obtained earlier in \cite{Vagnozzi:2017ovm}. 
In the $1\sigma$ range, the model has no points compatible with data in the NO case. In the IO case, the minimum values of both $m_{\text{lightest}}$ and $m_{ee}$ are around 0.03 eV. Given the $3\sigma$ ranges, both mass orderings are compatible with data. The minimum values of $m_{\text{lightest}}$ and $m_{ee}$ compatible with data are given by
\begin{eqnarray} 
(m_{\rm lightest})_{\rm min}^{\rm NO} \approx 0.025 ~{\rm eV}\,, &~&
(m_{ee})_{\rm min}^{\rm NO} \approx 0.021 ~{\rm eV} \,; \nonumber\\
(m_{\rm lightest})_{\rm min}^{\rm IO} \approx 0.026 ~{\rm eV}\,, &~&
(m_{ee})_{\rm min}^{\rm IO} \approx 0.029 ~{\rm eV}\,, 
\end{eqnarray}
respectively. Making use of the best cosmological constraint, $\sum m_i <0.12$ eV, we arrive at $m_{\rm lightest} < 0.31$ eV for NO and $< 0.17$ eV for IO. Most points in NO and all points in IO lie in this ``disfavoured'' region. On the other hand, few points lie in the ``very disfavoured'' region. 

\section{Conclusion}
\label{conclusion}

In this paper we have discussed a minimal model of trimaximal mixing in which the first column of the tri-bimaximal lepton mixing matrix is achieved via two modular $S_4$ groups, namely $S_4^l \times S_4^\nu$. The associated
moduli fields are assumed to be ``stabilised" at these two different symmetric points, where the misalignment leads to the lepton mixing. 
To be precise, one of these factors, $S_4^\nu$ 
acts in the heavy Majorana neutrino sector, under which the right-handed neutrinos transform as triplets, 
and is associated with a modulus field value $\tau_{SU}$ with residual $Z^{SU}_2$ symmetry.
The other factor $S_4^l $ acts in the Dirac charged lepton sector and is associated with a modulus field value $\tau_{T}$ with residual $Z^{T}_3$ symmetry. 

In addition there is a Higgs scalar $\Phi$ introduced to break the $S_4^l \times S_4^\nu$
down to a diagonal $S_4$ subgroup, yielding a Dirac neutrino Yukawa matrix proportional to $P_{23}$ at low energy (but above the seesaw scale).
The model here represents a simpler example of multiple modular symmetries than a previous model based on three modular symmetries, in which two Higgs 
scalars were required to break the three modular symmetries down to their diagonal subgroup.

In our chosen basis, the model leads to a diagonal charged lepton mass matrix, 
together with a heavy Majorana neutrino mass matrix which depends on three complex parameters, one
fewer than previous flavon models of TM$_1$ mixing which do not use modular symmetry at all. Together with the Dirac neutrino Yukawa matrix proportional to $P_{23}$,
this implies that the light effective left-handed neutrino Majorana mass matrix and the heavy Majorana mass matrix are diagonalised by the same unitary matrix,
namely the TM$_1$ lepton mixing matrix. The model therefore is subject to the usual TM$_1$ lepton mixing sum rules.

Apart from the usual predictions of TM$_1$ lepton mixing, the model also leads to a new neutrino mass sum rule,
which implies sizeable, quite degenerate, neutrino masses, with a marked preference for IO over NO. Much of the parameter space for the IO region
falls well inside the cosmologically disfavoured region. By contrast, some of the parameter space for the NO case falls
outside the cosmologically disfavoured region, with most points being not very disfavoured at the moment, 
although this conclusion could change with modest improvements in the cosmological limits.
In both IO and NO cases, the entire parameter space of the model can be probed by the planned neutrinoless double beta decay experiments.

In conclusion, we have proposed a minimal model of TM$_1$ lepton mixing based on having an independent modular $S_4$ symmetry acting 
in each of the charged lepton and neutrino sectors, respectively, where the two associated moduli respect different residual symmetries. 
The model, at the intermediate scale where only a single $S_4$ symmetry is conserved, does not involve any flavons but it does reply on a Higgs field breaking the two $S_4$ symmetries down to their diagonal subgroup.
The combination of the TM$_1$ lepton mixing sum rules and the new neutrino mass sum rule makes the proposed model
highly testable in the near future.


\section*{Acknowledgements}
SFK and YLZ acknowledge the STFC Consolidated Grant ST/L000296/1 and the European Union's Horizon 2020 Research and Innovation programme under Marie Sk\l{}odowska-Curie grant agreements Elusives ITN No.\ 674896 and InvisiblesPlus RISE No.\ 690575.
The authors also gratefully acknowledge the hospitality of Fermilab.


\appendix

\section{Group theory of $S_4$ \label{app:S4} } 

$S_4$ is the permutation group of 4 objects, see e.g.\ \cite{Escobar:2008vc,deMedeirosVarzielas:2017hen}.
The Kronecker products between different irreducible representations can be easily obtained:
\begin{eqnarray}
&\mathbf{1^{\prime}}\otimes\mathbf{1^{\prime}}=\mathbf{1}, ~~\mathbf{1^{\prime}}\otimes\mathbf{2}=\mathbf{2}, ~~\mathbf{1^{\prime}}\otimes\mathbf{3}=\mathbf{3^{\prime}}, ~~
\mathbf{1^{\prime}}\otimes\mathbf{3^{\prime}}=\mathbf{3},~~\mathbf{2}\otimes\mathbf{2}=\mathbf{1}\oplus\mathbf{1}^{\prime}\oplus\mathbf{2},\nonumber\\
&
\mathbf{2}\otimes\mathbf{3}=\mathbf{2}\otimes\mathbf{3^{\prime}}=\mathbf{3}\oplus\mathbf{3}^{\prime},~~
\mathbf{3}\otimes\mathbf{3}=\mathbf{3^{\prime}}\otimes\mathbf{3^{\prime}}=\mathbf{1}\oplus \mathbf{2}\oplus\mathbf{3}\oplus\mathbf{3^{\prime}},~~
\mathbf{3}\otimes\mathbf{3^{\prime}}=\mathbf{1^{\prime}}\oplus \mathbf{2}\oplus\mathbf{3}\oplus\mathbf{3^{\prime}} \,. \nonumber\\
\end{eqnarray}
%
\begin{table}[h!]
\begin{center}
\begin{tabular}{|c|ccc|}
\hline\hline
   & $\rho(T)$ & $\rho(S)$ & $\rho(U)$  \\\hline
$\mathbf{1}$ & 1 & 1 & 1 \\
$\mathbf{1^{\prime}}$ & 1 & 1 & $-1$ \\
$\mathbf{2}$ & 
$\left(
\begin{array}{cc}
 \omega  & 0 \\
 0 & \omega ^2 \\
\end{array}
\right)$ & 
$\left(
\begin{array}{cc}
 1 & 0 \\
 0 & 1 \\
\end{array}
\right)$ & 
$\left(
\begin{array}{cc}
 0 & 1 \\
 1 & 0 \\
\end{array}
\right)$ \\

$\mathbf{3}$ &  $\left(
\begin{array}{ccc}
 1 & 0 & 0 \\
 0 & \omega ^2 & 0 \\
 0 & 0 & \omega  \\
\end{array}
\right)$ &
$\frac{1}{3} \left(
\begin{array}{ccc}
 -1 & 2 & 2 \\
 2 & -1 & 2 \\
 2 & 2 & -1 \\
\end{array}
\right)$ &
$\left(
\begin{array}{ccc}
 1 & 0 & 0 \\
 0 & 0 & 1 \\
 0 & 1 & 0 \\
\end{array}
\right)$ \\

$\mathbf{3^{\prime}}$ &  $\left(
\begin{array}{ccc}
 1 & 0 & 0 \\
 0 & \omega ^2 & 0 \\
 0 & 0 & \omega  \\
\end{array}
\right)$ &
$\frac{1}{3} \left(
\begin{array}{ccc}
 -1 & 2 & 2 \\
 2 & -1 & 2 \\
 2 & 2 & -1 \\
\end{array}
\right)$ &
$-\left(
\begin{array}{ccc}
 1 & 0 & 0 \\
 0 & 0 & 1 \\
 0 & 1 & 0 \\
\end{array}
\right)$ \\ \hline\hline

\end{tabular}
\caption{\label{tab:rep_matrix_main} The representation matrices for the $S_4$ generators $T$, $S$ and $U$ used in the main text, where $\omega$ is the cube root of unit $\omega=e^{2\pi i/3}$.}
\end{center}
\end{table}
%

The generators of $S_4$ in the basis we used in the main text in different irreducible representations are listed in Table \ref{tab:rep_matrix_main}.  
This basis is widely used in the literature since the charged lepton mass matrix invariant under $T$ is diagonal in this basis. The products of $a \sim b \sim \mathbf{3}$ (or $a\sim b \sim \mathbf{3}'$) are expressed as
\begin{eqnarray}
(ab)_\mathbf{1} &=& a_1b_1 + a_2b_3 + a_3b_2 \,,\nonumber\\
(ab)_\mathbf{2} &=& (a_2b_2 + a_1b_3 + a_3b_1,~ a_3b_3 + a_1b_2 + a_2b_1)^T \,,\nonumber\\
(ab)_{\mathbf{3}} &=& (2a_1b_1-a_2b_3-a_3b_2, 2a_3b_3-a_1b_2-a_2b_1, 2a_2b_2-a_3b_1-a_1b_3)^T \,,\nonumber\\
(ab)_{\mathbf{3}'} &=& (a_2b_3-a_3b_2, a_1b_2-a_2b_1, a_3b_1-a_1b_3)^T \,.
\label{eq:CG2}
\end{eqnarray}
Here, $\mathbf{3}$ and $\mathbf{3}'$ represent the symmetric and antisymmetric triplet contractions, respectively \footnote{
Note that the difference of conventions for $\mathbf{3}$ and $\mathbf{3}'$ in this paper from those in e.g., \cite{King:2011zj} and \cite{Luhn:2013vna}, where $\mathbf{3}$ represents the antisymmetric triplet contraction for two $\mathbf{3}$ (or two $\mathbf{3}'$) and $\mathbf{3}'$ represents the symmetric triplet contraction. }. 
For $a \sim \mathbf{3}$ and $b \sim \mathbf{3}'$, the contractions are given by
\begin{eqnarray}
(ab)_{\mathbf{1}'} &=& a_1b_1 + a_2b_3 + a_3b_2 \,,\nonumber\\
(ab)_\mathbf{2} &=& (a_2b_2 + a_1b_3 + a_3b_1,~ -(a_3b_3 + a_1b_2 + a_2b_1))^T \,,\nonumber\\
(ab)_{\mathbf{3}'} &=& (2a_1b_1-a_2b_3-a_3b_2, 2a_3b_3-a_1b_2-a_2b_1, 2a_2b_2-a_3b_1-a_1b_3)^T \,,\nonumber\\
(ab)_{\mathbf{3}} &=& (a_2b_3-a_3b_2, a_1b_2-a_2b_1, a_3b_1-a_1b_3)^T \,.
\label{eq:CG2p}
\end{eqnarray}
The products of two doublets $a=(a_1, a_2)^T$ and $b=(b_1, b_2)^T$ are divided into
\begin{eqnarray}
(ab)_\mathbf{1} &=& a_1b_2 + a_2b_1 \,,\quad
(ab)_\mathbf{1^\prime} = a_1b_2 - a_2b_1 \,,\quad
(ab)_{\mathbf{2}} = (a_2b_2, a_1b_1)^T \,.
\label{eq:CG_doublets}
\end{eqnarray}

\end{document}